\documentclass[12pt]{article}
\usepackage[utf8]{inputenc}
\usepackage{graphicx} 
\usepackage{dcolumn} 
\usepackage{bm} 
\usepackage{amsfonts}
\usepackage{amsthm}
\usepackage{amsmath}
\usepackage{amssymb}
\usepackage{epsfig}
\usepackage{color}
\usepackage{textcomp}
\usepackage{hyperref}
\usepackage{titlesec}
\usepackage{slashed}
\usepackage{caption}
\usepackage{subcaption}
\usepackage{siunitx}
\usepackage{booktabs}
\usepackage{multirow}
\usepackage{cite}
\usepackage{placeins}
\usepackage{comment}
\usepackage{braket}
\usepackage{authblk}

\let\a=\alpha \let\b=\beta    \let\d=\delta \let\e=\varepsilon
  \let\h=\eta     
\let\m=\mu    \let\n=\nu             
\let\s=\sigma     
   \let\o=\omega

  \let\del=\nabla

\def\\{\hfill\break} \let\==\equiv

\let\0=\noindent

\let\dpr=\partial

\def\qed{\hfill\raise1pt\hbox{\vrule height5pt width5pt depth0pt}}

\def\be{\begin{equation}}
\def\ee{\end{equation}}
\def\bea{\begin{eqnarray}}\def\eea{\end{eqnarray}}

\begin{document}

\title{Autocatalysis of Graviton Production via the Gertsenshtein Effect for the Yang-Mills Field}

\author[1]{Andrea Palessandro \thanks{apalessandro@deloitte.com}}
\author[2]{Tony Rothman \thanks{tonyrothman@gmail.com}}
\affil[1]{\small Deloitte AI Institute}
\affil[2]{\small New York University, Department of Applied Physics (retired)}
\date{}

\maketitle

\begin{abstract}
\noindent As shown by Gertsenshtein in 1961, an external magnetic field can couple electromagnetic to gravitational waves, giving rise to oscillations between them. Viewed as a quantum process, the magnetic field has catalyzed the mixing of photon and graviton states. We show that the self-interaction terms of the Yang-Mills SU(2) field can take the place of the external magnetic field and autocatalyze the mixing of Yang-Mills bosons and gravitons. In the process, rotational symmetry is broken and the Yang-Mills boson acquires a mass.
\end{abstract}

\section{Introduction}
\setcounter{equation}{0}\label{intro}
\baselineskip 8mm

In a previous paper \cite{PR23}, we presented a straightforward derivation of the Gertsenshtein effect based on unpublished notes of Freeman Dyson \cite{Dyson05a}. As originally discovered by Gertsenshtein \cite{Gert61}, a large external magnetic field can catalyze the mixing of  electromagnetic and gravitational waves, resulting in a phenomenon reminiscent of neutrino oscillations. Essentially, the external field  couples the internal magnetic field of the electromagnetic wave to the spacetime curvature produced by the gravitational wave. The coupled system of equations for the electromagnetic and gravitational fields in a region where the  electromagnetic field is the only energy source admits oscillatory solutions.

Although the original Gertsenshtein mechanism is entirely classical, in quantum-mechanical language one can say that the external magnetic field has catalysed a resonant mixing between photon and graviton states. Such an interpretation, however, assumes that individual quanta of the gravitational field exist, a fact that has not, and may never be, experimentally established \cite{RB06, AP2020, Dyson05b}. 

If it is meaningful to talk about gravitational quanta, then despite the extremely weak coupling between the graviton and photon states, the exceptionally strong magnetic fields of neutron stars $\sim 10^{14}$ G make them plausible graviton factories. Such fields, however, are at approximately the Schwinger limit, where one expects electron-positron pairs to be created from the vacuum \cite{Zel73, Dyson05b}. Such pairs raise the local index of refraction, which in turn lowers the speed of light, destroying the presumed coherence between the electromagnetic and gravitational waves. Depending on the field strength, the effectiveness of the Gertsenshtein mechanism is diminished, if not  quenched altogether. That outcome should be investigated more closely. 

The ultimate drawback to the Gertsenshtein mechanism, however, is that it requires the imposition of an external magnetic field. In attempting to address this issue one notices that the original Yang-Mills SU(2) field, in contrast to the U(1) Maxwell field, contains self-interaction terms.  Evidently, the Yang-Mills (YM) field  might itself provide the  ``external magnetic field" required by the Gertsenshtein mechanism and consequently autocatalyze the mixing of graviton and boson states.  
 
 Given the importance of the original SU(2) YM field in the development of modern nonabelian gauge theories, it seems reasonable to examine the Gertsenshtein effect in that scenario, which is the purpose of this paper. Via a simple choice of potentials, we show that one of the ``magnetic field" components can be arranged to have a constant background value, providing the necessary catalysis for graviton-boson mixing. This field component breaks Lorentz (rotational) symmetry, in the process generating a mass term for the gauge boson even as it mixes with the gravitational field.
 
Our approach is step-for-step the same as the one we employed in \cite{PR23}  to derive the Gertsentein effect for the Maxwell field, and we recommend that the reader follow it side-by-side with that paper.  We write down the gravitational wave equation, as well as the YM wave equation and search for simultaneous, oscillatory solutions between gravitons and YM bosons.  As  mentioned above, we find that such solutions do exist for a simple choice of the potentials, establishing the autocatalysis of graviton production by the Gertsenshtein effect for the Yang-Mills field.
 
\section{Yang-Mills and Gravitational Wave Equations}
\setcounter{equation}{0}\label{YMfield}
 
To write down the wave equation for the Yang-Mills field, we first assume that the YM equations in flat space go over to the curved-space equations with the usual minimal substitutions, $ \text{comma} \rightarrow  \text{semicolon}$ and $\h_{\m\n}\rightarrow g_{\m\n}$. Then, the YM field equations become \cite{PS95,IZ}:
\be
\frac1{\sqrt{-g}}(\sqrt{-g} g^{\a\m}g^{\b\n} F^a_{\m\n})_{,\b} + q\e^{abc}g^{\a\m}g^{\b\n} A^b_\b F^{c}_{\m\n}= J^{a \alpha}. \label{YM}
\ee
Notice that the first term represents the ordinary Einstein-Maxwell equations, and is the only term required in deriving the electromagnetic Gertsenstein effect.  In the second term,  $q$ is the gauge coupling constant (not to be confused with the electric charge; taken to be dimensionless), $\e^{abc}$ is the antisymmetric permutation operator, and the $A^a$ are the three YM potentials, which will be designated by  Latin indices $a,b,c = 1...3$\footnote{We use units in which $G = c = 1$ and follow MTW \cite{MTW73} conventions throughout. In particular, the flat-space metric is taken to be $\h_{\m\n} = (-1,1,1,1)$. Greek indices $= 0...3$ are spacetime indices, while Latin indices $= 1...3$ are spatial indices. Repeated indices are always summed.}. Finally, $J^a$ is the YM four-current density.  

The YM field tensor is
\be
F^{a}_{\m\n} = \dpr_\m A^a_\n-\dpr_\n A^a_\m + q\e^{abc}A^b_\m A^c_\n. \label{YMF}
\ee
The energy-momentum tensor is the same as the ordinary electromagnetic energy-momentum tensor, except that one must trace over the internal indices:
\begin{equation}
    4 \pi T_{\mu \nu} =  F^a_{\mu \alpha} F^\alpha_{a\nu} - \frac{1}{4} g_{\mu \nu} F^a_{\alpha \beta} F_a^{\alpha \beta}.
\end{equation}
One can define ``magnetic" and ``electric" fields in exactly the same way as in the Maxwellian case, that is, $F^a_{ij} = \e_{ijk}B^a_k$ and $F^a_{i0} = E^a_i$. We follow this convention; however, it should be understood that these are not the Maxwellian electric and magnetic fields. With these designations, the stress-energy tensor for flat space reads:
\be
4\pi T_{ij}= -(E^a_iE^a_j + B^a_iB^a_j)+\frac1{2}\d_{ij}(\bf E^a \cdot E^a + B^a \cdot B^a).\label{YMT}
\ee

Next, we assume that the full spacetime metric is a small perturbation of the Minkowski background, such that $g_{\mu \nu} = \eta_{\mu \nu} + h_{\mu \nu}$, with $h_{\mu \nu} \ll 1$, and we carry out our calculations in the Transverse Traceless (TT), or Lorenz, gauge\footnote{After Ludvig Lorenz, the Danish physicist, 1829-1891.}. Apart from the  definition of $T_{\mu \nu}$, the (linearized) gravitational wave equation retains its usual form:
\be
\Box h_{\mu\nu} = -{16\pi} T_{\mu \nu},\label{boxh}
\ee
with $\Box \equiv \dpr^\m \dpr_\m$. As in \cite{PR23}, we ignore gravitational backreaction, which is a good approximation when $h_{\mu \nu} \ll 1$, and take the energy momentum tensor to be entirely that of the electromagnetic field.

Then, working to first order in $h_{\mu \nu}$, (\ref{YM}) yields
\begin{equation}
    \begin{split}
        \h^{\a\m} \h^{\b\n} F^a_{\m\n},_\b - \h^{\a\m}h^{\b\n}F^a_{\m\n},_\b -\h^{\a\m}h^{\b\n},_\b F^a_{\m\n} -\h^{\b\n}h^{\a\m}F^a_{\m\n},_\b -\h^{\b\n}h^{\a\m},_\b F^a_{\m\n}\\
        + q\h^{\a\m}\h^{\b\n}\e^{abc}A^b_\b F^{c}_{\m\n} - q( \h^{\b\n}h^{\a\m} + \h^{\a\m}h^{\b\n})\e^{abc}A^b_\b F^{c}_{\m\n}  = J^{a \alpha}.
    \end{split}
\end{equation}
In the Lorenz gauge $h^{\b\n},_\b \equiv 0$. Discarding terms $\sim h F$ as small but keeping terms $\sim \partial h F$ as potentially large leaves
\begin{equation}
    \h^{\a\m}\h^{\b\n}F^a_{\m\n},_\b -\h^{\b\n}h^{\a\m},_\b F^a_{\m\n} + q\h^{\a\m}\h^{\b\n}\e^{abc}A^b_\b F^c_{\m\n} = J^{a \alpha}. \label{YM2}
\end{equation}
For $\a = i$, (\ref{YM2}) becomes
\be\label{YM2a}
-\dot E^a_i + \e_{i j k}B^a_{k,j} + \dot h_{ij}E^a_j - h_{ij,k}\e_{jk\ell}B^a_\ell
-q\e^{abc}A^b_0 E^c_i + q\e^{abc}\e_{i j k} A^b_j B^c_k = J^a_i.
\ee
As in case of the Maxwell field, to  transform this expression into a wave equation we take its curl to get:
\begin{equation}\label{curlYM2}
    \begin{split}
        - \e_{i j k} \dot{E}^a_{k,j} + B^a_{j,ji}-B^a_{i,jj} + \e_{ijk}(\dot h_{kp}E^a_p)_{,j} \\
        - \e_{i p r} \e_{j k\ell}(h_{rj,k} B^a_\ell)_{,p} - q\e_{ijk}\e^{abc}(A^b_0 E^c_k)_{,j} \\
        + q\e^{abc}[( A^b_iB^c_j)_{,j}-(A^{b}_jB^c_i)_{,j}]= \e_{ijk} J^a_{k,j}.
    \end{split}
\end{equation}

In contrast to the Maxwellian case, for the YM field we cannot automatically set $\bf \del \cdot B = 0 $ or $\bf \del \times E = -\dpr B/\dpr t$ because the  homogeneous YM equations differ from the homogeneous Mawell equations. We will, however, attempt to eliminate the first and second terms above. The homogeneous YM equations (Bianchi identities) are
\be
(\dpr_\m F^a_{\n\s} + q\e^{abc}A^b_\m F^c_{\n\s}) + \mathrm{cyclical \  permutations \ of \ \m\n\s} = 0,
\ee
where we work in the so-called adjoint representation. As in the Maxwellian case, all indices must be distinct.  For $(\m,\n,\sigma) = (0,j,k)$, and upon multiplication by $\e_{ijk}$ one finds
\be
\dot B^a_i - \e_{ijk}E^a_{j,k} + q\e^{abc}A^b_0B^c_i + q\e_{ijk}\e^{abc}A^b_jE^c_k = 0. \label{YMh0}
\ee
The first two terms are the Maxwellian ones. For $(\m, \n, \sigma) = (i,j,k)$, and upon multiplication by $\e_{ijk}$, one has
\be
B^a_{j,j} + q\e^{abc}A^b_j B^c_j = 0,\label{YMhi}
\ee
which is the covariant generalization of $\bf \del \cdot B = 0$.  Taking the time derivative of (\ref{YMh0}) and the gradient of (\ref{YMhi}) to eliminate the first two terms in (\ref{curlYM2}) yields
\begin{equation}\label{YM3}
    \begin{split}
        \ddot{B}_i^a -B^a_{i,jj} + \e_{ijk}(\dot h_{kp}E^a_p)_{,j} - \e_{i p r} \e_{j k\ell}(h_{rj,k} B^a_\ell)_{,p} \\
        - q\e_{ijk}\e^{abc}(A^b_0 E^c_k)_{,j} + q\e^{abc} (A^b_0B^c_i)_{,0} + q\e_{ijk}\e^{abc}(A^b_jE^c_k)_{,0} \\
        + q\e^{abc}[( A^b_iB^c_j)_{,j}-(A^{b}_jB^c_i)_{,j}] - q\e^{abc}(A^b_j B^c_j)_{,i}= \e_{ijk} J^a_{k,j}.
    \end{split}
\end{equation}
The first four terms are identical to those obtained in the Maxwellian case, as they must be.  However, this is a nonlinear equation with multiple solutions and it is not immediately obvious that it admits wave solutions. Nevertheless, in the case of the YM field alone (without the gravitational interaction terms), it has been known for some time that it does \cite{Coleman77,ML78,KL79,BH84}.

\section{Oscillatory Solution}
\setcounter{equation}{0}\label{Solutions}

We search for the simplest possible simultaneous solution of the gravitational and YM wave equations. As in \cite{PR23}, we assume that the gravitational wave is propagating in the $z$-direction, $h_{\mu \nu}(z,t)$, with the two independent polarizations being $h_{11}=-h_{22}$, and $h_{12} = h_{21}$. In the following, we use dots for time derivatives, and primes for $z$-derivatives.  

Note that if all the potentials $A^b$ are the same, the coupling terms in (\ref{YM3}) evidently disappear due to the symmetry of the permutation operator. For the choice of potentials
\begin{equation}
    \begin{split}
        A^1 &= (0, f(z,t), 0, 0)\\
        A^2 &= (0, 0, A, 0)\\
        A^3 &=  (0, 0, 0, A),
    \end{split}
\end{equation}
where $A =$ constant, one finds the electric and magnetic fields from (\ref{YMF}) to be
\be
E^1_1 = -\dot f \quad ; \quad B^1_1 = + q A^2 \quad ; \quad B^1_2 = + f' \quad ; \quad B^2_2 = + qAf \quad ; \quad B^3_3 = + qAf .
\ee
These potentials were chosen in order to make $B^1_1 = q A^2= \text{constant}$. This field component thus acts as the ``external" magnetic field of the electromagnetic case, while $E_1^1$ and $B_2^1$ act as  the electric and magnetic fields of the electromagnetic wave, respectively. 

With this choice of potentials, and taking the homogeneous case $J^1 = 0$, (\ref{YM3}) gives for $a = 1, i = 2$\footnote{All other equations in (\ref{YM3}) are trivially satisfied with our choice of potentials and $J^a_k = 0$, except for the combinations $(a=2,i=2)$ and $(a=2, i=1)$, which give the additional constraints $qA \left[ (h'_{11} f)' - 2f'' \right] = J^2_{1,3}$, and $qA \left[(h'_{12}f)' + qAf'\right] = J^2_{2,3}$, respectively. By assumption $h_{ij} \ll 1$ and $h'_{ij} \ll q A^2$, since $(h'_{ij})^2 \ll q^2A^4 = (B^1_1)^2$, the dominant energy component of the system.  Thus with $J^2_\mu \equiv J^2_\mu (z,t)$, the two constraints can be satisfied to leading order by choosing $J^2_1 = - 2 qA f'$ and $J^2_2 = q^2A^2f$. One is then free to choose the other two components $J_0^2$ and $J^2_3$ such that $J^2_{0,0} + J^2_{3,3} = 0$ to ensure four-current conservation, as these give vanishing contributions to the right-hand side of (\ref{YM3}).}:
\be\label{B12}
\ddot B^1_2 - (B^1_2)'' + (\dot h_{11} E^1_1)' + (h_{11}' B^1_2)' - (h'_{12}B^1_1)' + qA[(B^3_3)' + (B^2_2)'] = 0.
\ee
The gravitational wave equation (\ref{boxh}) for the $h_{12}$ polarization is
\begin{equation}
    - \ddot{h}_{12} + h_{12}'' = {4}  (E_1^a E_2^a + B_1^a B_2^a) = 4 q {A^2} B_2^1.
\end{equation}
Assuming that $B_1^1$ is much larger than the electromagnetic wave components $E_1^1$ and $B_2^1$, and writing $B_2^1 = f' \equiv b$ and $h_{12} \equiv h$, one arrives at the coupled equations:
\be\label{ymh12}
\ddot b - b'' + 2 q^2 A^2 b= q A^2 h'' ,
\ee
and
\be\label{gwh12}
\ddot h - h'' = -4 qA^2 b.
\ee
The above are the final equations in \cite{PR23} except for the $2q^2A^2b$ term in (\ref{ymh12}) and that the background magnetic field $B_0$ is  replaced  by $qA^2$. Note that the constant background magnetic field $B_1^1$ breaks Lorentz invariance by picking out a preferred direction in space, and generates a mass term for the boson field $B^1_2$ equal to $m^2 \equiv 2qB_1^1$.  To an extent this reminds one of the  Higgs mechanism, although it is not ``spontaneous'' symmetry breaking; rather, symmetry is broken explicitly by arranging for one magnetic field component to be larger than the others.

Given that the equations are essentially those as in \cite{PR23} (plus a mass term), we assume that both  $b$ and $h$ can be decomposed into  products of a rapidly varying function and a slowly varying function of the form
\begin{equation}\label{hbdef}
    h = \mathcal{A} e^{i\omega_r(z-t)} e^{-i\omega_s(z+t)} \ \ ; \ \ b = \mathcal{B} e^{i\omega_r(z-t)} e^{-i\omega_s(z+t)},
\end{equation}
with $\omega_r \gg \omega_s$. The rapidly varying components define the field (or particle) energy $\omega_r$, while the slowly varying components describe the oscillation between boson and gravitational states with frequency $\omega_s$.\footnote{Note that in (\ref{hbdef}) $h \propto b$, therefore no coherent oscillations can occur between these two fields. The reason is that, as in \cite{PR23}, the actual gravitational state that mixes with $b$ is $\Tilde{h} \equiv \dot{h}/2$, as this is the field that enters into the gravitational energy density. The factor of $i$ coming from the differentiation with respect to time rotates the solution by $\pi/2$, giving the required phase difference between the two fields.} 

In writing down the decomposition (\ref{hbdef}) we have assumed $\omega_r \gg m \sim qA$, so that $b$ can be written, like $h$, as a (nearly) massless wave packet\footnote{We cannot actually set $m = 0$, since in the present case the same mechanism gives rise to both mass \textit{and} the catalyzing magnetic field. That is, letting $m \to 0$ requires  $A \rightarrow 0$, which in turn implies $B_1^1 \rightarrow 0$.  In that limit the mixing length  becomes infinite, since there is no coupling between the YM boson and the graviton.}. As we show below, given $\omega_r \gg q A$, the condition $\omega_r \gg \omega_s$ is always satisfied, so our ansatz is fully consistent. We will also see that, in contrast to the electromagnetic case, the presence of a small but non-zero bosonic mass makes the mixing length explicitly dependent on the wave frequency. 

Inserting  ansatz (\ref{hbdef}) into (\ref{ymh12}) and (\ref{gwh12}) yields
\begin{equation}
(4\omega_r\omega_s  - 2 q^2 A^2) b= q A^2 (\omega_r-\omega_s)^2 h,
\end{equation}
and
\begin{equation}
    \omega_r \omega_s h= q A^2 b.
\end{equation}
Taking $(\o_r-\o_s)^2 \approx \o_r^2$, together these can be easily solved for $\omega_s$ to obtain (choosing the positive solution):
\be\label{omegas}
\o_s = \frac{q^2A^2}{4 \omega_r} \left( 1 + \sqrt{1 + \frac{4 \omega_r^2}{q^2}}\right).
\ee
Thus, as stated, the mixing frequency $\omega_s$ now depends on  $\omega_r$. Furthermore, $\omega_r \gg \omega_s$ is automatically satisfied whenever $\omega_r \gg q A$.

In the high-energy limit $\omega_r \gg q$, (\ref{omegas}) becomes
\begin{equation}\label{omegasel}
    \omega_s \equiv \frac{1}{L} = \frac{qA^2}{2},
\end{equation}
which is the electromagnetic result of \cite{PR23}, given that there $B_0 = q A^2$. When $A=1$, the condition $\omega_r \gg q A$ reduces to $\omega_r \gg q$, meaning that letting $A \to 1$ is equivalent to taking the high-energy limit.\footnote{The linearized field equations are  valid only as long as the energy density in the fields is subplanckian. Given that by assumption $B_0$ dominates the energy density of the coupled system, that condition becomes $B_0^2 \sim q^2A^4 \ll 1$. This is always true when $q \ll 1$, even for $A=1$.  Recall also that $h \ll 1$ has been imposed in the derivation of the linearized Einstein equations (\ref{boxh}). As a consequence, the gravitational wave amplitude in (\ref{hbdef})  must obey the condition $\mathcal{A} \ll 1$.} 

One can prove our results in the high-energy limit more simply by the identical procedure employed in \cite{PR23}. Assume that both $b$ and $h$ can be decomposed into  products of a rapidly varying function and a slowly varying function of the form $f = f_r(z-t)f_s(z+t)$. In these units $f'_r = -\dot f_r$, but $f'_s = +\dot f_s$, and by assumption $\dot f_r \gg \dot f_s$.  

Then, substituting $b = b_r(z-t)b_s(z+t)$ and $h = h_r(z-t)h_s(z+t)$ into (\ref{ymh12}) and (\ref{gwh12}) yields
\begin{equation}\label{brbs}
    -4 b_r'b_s' + 2 q^2 A^2 b_r b_s = q A^2 (h_r''h_s + 2 h_r'h_s'+h_r h_s''),
\end{equation}
and
\begin{equation}\label{hrhs}
    h_r'h_s' = q {A^2} b_r b_s.
\end{equation}
Now substituting (\ref{hrhs}) for the mass term in (\ref{brbs}) gives
\be\label{brbs2}
 -4 b_r'b_s' + 2 q   h'_r h'_s = q A^2 (h_r''h_s + 2 h_r'h_s'+h_r h_s'').
 \ee
In the case $A = 1$, the mass term identically cancels the second term on the right and only the last term in (\ref{brbs2}) need be dropped as small compared to the others (in fact a better approximation than the one used in \cite{PR23}).

Thus the equations can now be solved by inspection.  Pick
\be\label{hbeq}
h'_r = \frac{\sqrt q}{a} b_r \ \ ; \ \ h'_s = a\sqrt{q}b_s  \ \ ; \ \ b'_s=-\frac{q^{3/2}}{4a}  h_s,
\ee
with $a$ an arbitrary constant.  

Letting $a=1/2$, the  solution for the slowly varying functions is
\be
h_s = {\cal A}\sin\left(\frac{z+t+\phi}{L}\right)\ \  ;\ \ b_s = {\cal A} \sqrt{q} \cos\left(\frac{z+t+\phi}{L}\right),
\label{GEymsol}
\ee
where the mixing length is now
\begin{equation}
    L = \frac{2}{q},
\end{equation}
which is the same as (\ref{omegasel}) for $A=1$.

\section{Discussion}
We have shown that a simple choice of YM potentials can produce a mixing between the YM and gravitational fields. If the gravitational field is quantized, the mixing gives rise at the quantum level to oscillations between YM bosons and gravitons. 

The particular configuration of potentials we adopt breaks rotational symmetry by selecting a preferred direction in space, which corresponds to that of the constant background ``magnetic" field $B_1^1$. This in turn gives rise through self-interaction terms to a mass for the mixing boson proportional to $2 q B_1^1$, and provides the required coupling between boson and graviton states.

Of course, there are an infinite number of choices of  potentials, most of which will certainly not admit oscillatory solutions.  We have also made several simplifying assumptions about the size of the field components and it would be desirable to relax these to solve the full set of equations, if necessary numerically.  Nevertheless, we have given a ``proof of concept" that if the Yang-Mills SU(2) field exists in nature, it has the potential to autocatalyze the production of gravitons. 

A phenomenon that appears closely related to this result is the parametric resonance in the SU(2) gauge field resulting from spacetime oscillations, which has been recently investigated in \cite{Dave23}.  There, the authors consider the effect of varying the amplitude and frequency of the metric oscillations $h$ on the SU(2) potentials $A^a$.  Although they assume a sinusoidal form for $h$ and do not search for wave-packet solutions with the SU(2) B-field, oscillations of $h$ are fed into the gauge potentials via their covariant spacetime derivatives.  Numerical evolution shows a resonant growth of the Fourier momentum modes of $A^a$, which can produce exponential growth in certain physical quantities, such as $\bf E\cdot B$.  This behavior, while not identical to the one we have discussed, does seem compatible.

The Gertsenshtein mechanism for the SU(2) field may also have implications for the study of the early universe. In \cite{PR23} we showed that the Gertsenshtein mechanism and axion-photon mixing are closely analogous, in fact almost identical effects. Recently, inflationary models have been proposed in which the axion is coupled to SU(2) fields, much as we have done here between the SU(2) field and gravitons. In \cite{Adshead:2012kp}, for example, an oscillating axion potential causes inflation rather than the gauge potential; nevertheless, it is possible to imagine a scenario in which our dominant gauge potentials induce the de Sitter phase, even as they generate particles by some variant of the Gertsenshtein effect.   

At the least, it has been known for some time that neutrinos and gauge bosons undergo mixing; axion searches presume that axion-photon oscillations exist as well. This paper's result extends the list of particles that may undergo inter-species mixing, and  we may guess that the phenomenon will assume ever more importance in the study of the early universe.\\

\noindent {\bf Acknowledgements}

\noindent We would like to thank the anonymous referees for their constructive comments, which helped to improve the paper.

\newpage

{\small

\end{document}